\documentclass[12pt]{article}

\usepackage{amsmath,amssymb,graphicx} 
\usepackage{epsf}
\usepackage{pstricks}
\usepackage{cite}
\newcommand{\drawsquare}[2]{\hbox{%
\rule{#2pt}{#1pt}\hskip-#2pt
\rule{#1pt}{#2pt}\hskip-#1pt
\rule[#1pt]{#1pt}{#2pt}}\rule[#1pt]{#2pt}{#2pt}\hskip-#2pt
\rule{#2pt}{#1pt}}
\newcommand{\Yfund}{\drawsquare{7}{0.6}}
\newcommand{\Ysymm}{\drawsquare{7}{0.6}\hskip-0.6pt%
    \drawsquare{7}{0.6}}

\newcommand{\beq}{\begin{eqnarray}}
\newcommand{\eeq}{\end{eqnarray}}

\newcommand{\centeron}[2]{{\setbox0=\hbox{#1}\setbox1=\hbox{#2}\ifdim

\wd1>\wd0\kern.5\wd1\kern-.5\wd0\fi \copy0

\kern-.5\wd0\kern-.5\wd1\copy1\ifdim\wd0>\wd1
                                       \kern.5\wd0\kern-.5\wd1\fi}}
\newcommand{\ltap}{\>\centeron{\raise.35ex\hbox{$<$}}
                               {\lower.65ex\hbox{$\sim$}}\>}
\newcommand{\gtap}{\>\centeron{\raise.35ex\hbox{$>$}}
                               {\lower.65ex\hbox{$\sim$}}\>}

\newcommand\ZZ{\hbox{\zfont Z\kern-.4emZ}}
\font\zfont = cmss10 

\begin{document}
\begin{titlepage}

\vskip1.5cm
\begin{center}
{\huge \bf A GIM Mechanism}
\vskip.1cm
\end{center}
\begin{center}
{\huge \bf from Extra Dimensions}

\vskip.1cm
\end{center}
\vskip0.2cm

\begin{center}
{\bf Giacomo Cacciapaglia$^a$, Csaba Cs\'aki$^b$, Jamison
Galloway$^a$, Guido Marandella$^a$, John Terning$^a$ and Andreas
Weiler$^b$}

\end{center}
\vskip 8pt

\begin{center}
$^{a}$ {\it Department of Physics, University of California, Davis,
CA
95616.} \\
$^{b}$ {\it Institute for High Energy Phenomenology\\
Newman Laboratory of Elementary Particle Physics\\
Cornell University, Ithaca, NY 14853, USA } \\
\vspace*{0.3cm}
\end{center}

\vglue 0.3truecm

\begin{abstract}
\vskip 3pt \noindent
We explore how to protect extra dimensional models from large flavor
changing neutral currents by using bulk and brane flavor
symmetries. We show that a GIM mechanism can be built in to warped space 
models such as Randall-Sundrum  or composite Higgs models if 
flavor mixing is introduced via UV brane kinetic mixings for right handed 
quarks. We give a realistic implementation both for a model with minimal 
flavor violation and one with next-to-minimal flavor violation. 
The latter does not suffer from a CP problem. We consider some of the existing experimental
constraints on these models implied by  precision electroweak tests. 

\end{abstract}

\end{titlepage}



\section{Introduction}
\label{sec:Intro} \setcounter{equation}{0} \setcounter{footnote}{0}

Extra dimensional theories offer new avenues for flavor physics.
Following the suggestion of Arkani-Hamed and Schmaltz~\cite{AS},
the standard approach is to use the overlaps of wave
functions in extra dimensions to generate the fermion mass
hierarchy. Since the fermions are physically located in different places this is referred to as the
split-fermion  approach. If implemented in Randall-Sundrum-type~\cite{RS1} warped space (as suggested
in~\cite{warpedflavor,HuberShafi}) it has the added benefit that unwanted
4-fermi operators leading to flavor
changing neutral currents (FCNC's) are suppressed by a high UV
scale (assuming that the fermions are mostly localized around the UV
brane). The main advantage of this approach is that it generates the
fermion mass hierarchy without flavor symmetries. The price to pay
for the absence of the flavor symmetries is that the structure of quark
mixing is much more complex than in the usual Cabibbo--Kobayashi--Maskawa
(CKM) picture \cite{CKM}, in
particular one would expect more mixing angles and CP phases to be
physical~\cite{Gustavo,warpedconstraints}. 
However, after many years of  running B-factories
the experimental results seem to suggest that flavor mixing is well
described by the standard CKM picture with a single CP violating
phase as in the standard model (SM). For the RS model these constraints would
imply that the mass of the lightest KK gauge bosons would have to be of order
$\sim 8$ TeV for generic choices of the parameters~\cite{NMFVbound}.
Given these constraints one should revisit the question
of how to introduce flavor physics in extra dimensional scenarios that
will reproduce CKM. An added motivation for this study comes from
the anti-de Sitter/conformal field theory (AdS/CFT) correspondence~\cite{AdSCFT,AdSCFTpheno}. 
We have learned over the past few years
that some simple extra dimensional setups behave like
weakly coupled duals of approximately conformal (walking)  technicolor theories~\cite{higgsless}.
Technicolor models are notorious for their problems with FCNC's and
the simplest known technicolor models incorporating a Glashow-Illiopoulos-Maiani (GIM)
mechanism \cite{GIM} are terrifyingly complex even for trained model builders~\cite{tcgim}.

In this paper we suggest an alternative general approach to flavor
physics in extra dimensions based on flavor symmetries. This
approach can be applied to warped space models including traditional Randall-Sundrum
1 (RS1), composite Higgs models or higgsless models, and also to flat
extra dimensional models (for example for gauge-Higgs unification models
in flat space). 
The main feature of this construction is that the bulk has a large flavor symmetry, while the IR brane where the
SM Yukawa couplings are localized still preserves a large diagonal
subgroup of the symmetry. Flavor mixing is
then introduced via kinetic mixing terms of the right-handed (R) fields on the UV brane. Therefore, higher 
dimensional flavor violating operators are forbidden by the flavor symmetries everywhere but on the UV brane. However 
the suppression scale for flavor violating 4 fermi operators on the UV brane is very large, so higher dimensional
operators will never pose a flavor problem (contrary for example to UED models where there is a GIM mechanism for the lowest dimensional
operators~\cite{UEDGIM} but there is no suppression mechanism against flavor violation from higher dimensional operators).

We
will show that this approach can incorporate a  GIM
mechanism and reproduce
the SM CKM picture. Thus it can be viewed as the
simplest implementation of minimal flavor violation\footnote{Which posits that that all flavor violation and the CKM matrix arise from the same source.} (MFV)~\cite{MFV}
 in extra dimensional theories. The
downside of these constructions is that we are no longer trying to
{\it explain} the fermion mass hierarchy, rather we want to
accommodate it with the least amount of flavor structure.

In order to find a realistic implementation of this idea we also need 
to make sure that the traditional  precision electroweak bounds are satisfied
beyond the flavor constraints. Incorporating a heavy top quark will
require us to slightly modify the simplest toy example, by making 
sure that the large top quark mass does not feed into the electroweak 
precision constraints of the light quarks (but still leaving sufficient 
flavor symmetries). This can be achieved by using the 
modified representations under the custodial symmetry for the quarks~\cite{newcustodial}.
We present two examples of this sort, one an example of minimal flavor 
violation (MFV)~\cite{MFV}, and one of next-to-minimal flavor violation (NMFV)~\cite{NMFV} with no CP problem.

This paper is organized as follows. In Sec.~2 we introduce the basic construction for 
flavor mixing via UV brane kinetic terms. We use symmetry arguments to show that 
such a setup indeed has a GIM mechanism built in, and then also explain the origin
of the GIM mechanism in  5D language using wave function orthogonality. We also 
show how the CKM matrix emerges in this picture. In Sec.~3 we give the CFT interpretation
of this setup, and also consider some of the bounds on the extra gauge bosons. In Sec.~4 we
show how a realistic model incorporating a heavy top can be obtained. We focus on models 
with a light Higgs on the IR brane (such as the RS1 model~\cite{RS1,ADMS} or models with a composite Higgs~\cite{MCH})
and a Kaluza-Klein (KK) mass scale $\gtrsim 3$ TeV. The first model we present has a full GIM mechanism built in, and
we check explicitly that it can also be consistent with the other  precision electroweak
bounds. The second model is an implementation of next-to-minimal flavor violation (NMFV) (by which we mean that all additional flavor violation has to go through the
third generation), but we argue that it does not have a CP problem. We conclude in Sec.~5.

\section{Symmetry considerations}
\label{sec:symmetry} \setcounter{equation}{0}
\setcounter{footnote}{0}

We will be using the following setup: there will be an exact flavor
symmetry in the bulk separately for left-handed (L) and right-handed fermions\footnote{We will use L to describe the 5D Dirac fermion field that contains a Weyl (chiral) left-handed zero-mode in the absence of  Higgs Yukawa couplings, and similarily with R for right-handed. When we need to distinguish between left and right-handed components of the Dirac fermion we will use $\chi$ and $\psi$ respectively.}.
To be concrete, in this section we will consider three generations
of quarks, and we will leave to the reader the straighforward
extension to leptons. In order to incorporate a custodial symmetry
necessary to be in agreement with  precision electroweak tests we
assume that the bulk electroweak gauge group is
SU(2)$_L\times$SU(2)$_R\times$U(1)$_{X}$, broken to the SM gauge
group by boundary conditions on the UV brane,
SU(2)$_R\times$U(1)$_X\to$U(1)$_Y$. It will be further broken to 
U(1)$_{em}$ on the IR brane, either by higgsing or boundary
conditions.  The largest global flavor
symmetry that we can impose in the bulk is a
U(3)$_Q\times$U(3)$_{uR}\times$U(3)$_{dR}$: this would be broken to
U(3)$_Q\times$U(3)$_{qR}$ by gauge interactions if the right-handed
quarks are embedded in the same multiplet of SU(2)$_R$. This
symmetry implies for example that the bulk masses of the fermions
with a given quantum number are precisely equal and that there is no
kinetic mixing. On the IR brane we assume that the quark Yukawa couplings
(or mass terms in Higgsless scenarios)
break the flavor symmetry to the family symmetry U(3)$_V$, the vector (diagonal) subgroup of the
three U(3)'s. This is achieved by flavor independent Yukawa couplings proportional to the identity
matrix in family space for both the
up-type and down-type quarks.\footnote{Note, that if we started with
a single right handed bulk multiplet for up and down at this stage
the unbroken flavor symmetry would be identical.} Finally, both the
splitting among the quark masses of different generations and flavor
mixing are achieved via kinetic mixing terms among the right handed
up-type and down-type quarks on the UV-brane where the gauge group
is broken to SU(2)$_L\times$ U(1)$_Y$, thus allowing kinetic
mixing that distinguish between up and down. The UV mixing terms will generically break the flavor
symmetry U(3)$_u\times$U(3)$_d\to$U(1)$_{uR}\times$U(1)$_{dR}$, where all R up-type quarks transform by the same phase under U(1)$_{uR}$. Such
a breaking pattern implies that, in order to avoid massless Goldstone
bosons or massless flavor gauge bosons in the bulk, the largest
flavor group that we can have in the bulk is
SU(3)$_Q\times$SU(3)$_{qR}$, independently of the fermion
representations. Thus if $uR$ and $dR$ are in different representations there must be
an additional source of breaking in the bulk to reduce the symmetry to SU(3)$_{qR}$.

We will now argue that this setup, together with the assumption that
only two kinetic mixing terms are allowed on the UV brane for up and
down quarks, in fact results in an extra dimensional GIM mechanism,
that is there will be no tree-level FCNC's generated, and  MFV via the
CKM matrix only.

To show this,
let us first turn off the charged current interactions. In the
neutral current sector in the bulk we have a bigger global symmetry,
U(3)$_{uL}
\times$U(3)$_{dL}\times$U(3)$_{uR}\times$U(3)$_{dR}$.  This
is due to the fact that neutral currents can not mix and up- and
down-type quarks.
On the IR brane the Yukawa couplings break the symmetry to the
vector (diagonal)  subgroup U(3)$_{uV}\times$U(3)$_{dV}$. So these are
the symmetries we can use to diagonalize the kinetic terms on the
UV brane. If we only have a kinetic mixing in the R quarks on
the UV brane, then we can use the
SU(3)$_{uV}\times$SU(3)$_{dV}$ symmetry to diagonalize the
kinetic terms. Note, that the extra U(1) factors leave the kinetic
mixing terms invariant, so they cannot be used to  further simplify
the kinetic matrix. While these kinetic terms will be diagonal, they
will not be proportional to the unit matrix, and so the diagonal
(non-equal) components clearly break the
U(3)$_u\times$U(3)$_d$ symmetry to
U(1)$_u\times$U(1)$_c\times$U(1)$_t\times$U(1)$_d\times$U(1)$_s\times$U(1)$_b$.
This symmetry is sufficient to eliminate FCNC's not just via the ordinary
$Z$-boson, but also through the $Z^\prime, g^\prime$, etc. KK modes. However,
non-universalities in the diagonal couplings will still be
generated at higher order in quark masses.

If there are additional kinetic mixings, for instance for the L
fields, on the UV brane, then (unless the kinetic matrices for the L
and R fields are simultaneously diagonalized) one will break the
flavor symmetry completely, and there will be FCNC's. This may also
happen if  the multiplets containing the right-handed quarks also
contain extra exotic quarks, made heavy via boudary conditions, that
mix with the SM ones via the Yukawa interactions (IR mass terms).

In the charged current (CC) sector there are no residual
diagonal U(1) symmetries, leading to the possibility of flavor
mixing in CC's. The reason behind the flavor violation in this setup
is that the U(1) flavor symmetries are misaligned in the bulk and
on the UV brane due to the kinetic mixing terms for the R
fermions. In fact, one needs to do a different rotation on the up and down-type
fields, which is not an invariance of the bulk.
This misalignment becomes physical in the CC interactions.

The origin of the GIM mechanism and the emergence of the 
CKM matrix can be seen explicitly if we consider the bulk fields
$\chi_L$, an SU(2)$_L$ doublet that contains the left-handed
doublet, and $\psi_R^{u,d}$, the fields containing the right-handed
quarks. As already mentioned, it is not crucial which representation
of SU(2)$_R$ $\psi_R^{u,d}$ are embedded in, since this symmetry is
broken on the UV brane where the relevant flavor mixings are
introduced. In the following, we will use $\chi$ and $\psi$ to
indicate the left and right-handed helicity components of the bulk fields. On
the UV brane, we can write two kinetic terms of the form:

\beq \mathcal{L}_{UV} =  \left. i \psi_R^\alpha\,  \sigma_\mu
D^\mu\,  \mathcal{K}^{\alpha \beta} \bar \psi_R^\beta
\right|_{z=z_{UV}}\,, \eeq both for up and down quarks, with
different mixing matrices $\mathcal{K}_u$ and $\mathcal{K}_d$. To
simplify the notation, we will suppress the weak isospin index as
the following discussion can be separately applied both to up and
down quarks. $\mathcal{L}_{UV}$ determines the BC's on the UV brane
for the R fields:

\beq \label{eq:BCUV} \chi_R (z_{UV}) = m\, \mathcal{K}\cdot \psi_R
(z_{UV})\,. \eeq The key point is that this is the {\it only} source
of flavor mixing: in fact both the bulk equation of motion and the
remaining boundary conditions are flavor diagonal.
 We can therefore
solve the equations of motion for all the fields and impose the IR
BC's  (and remaining BC's on the UV brane): this is enough to
determine uniquely the wave functions up to an overall
normalization. The solutions will look like:

\beq \label{eq:wf}
\begin{array}{cc}
\chi_{L}^\alpha = A^\alpha f_L (m, z)\, ,  & \chi_{R}^\alpha = A^\alpha f_R (m, z)\, , \\
\psi_{L}^\alpha = A^\alpha g_L (m, z)\, ,& \psi_{R}^\alpha = A^\alpha
g_R (m, z)\, .
\end{array} \eeq
It is crucial here that the functions $f_{L,R}$ and $g_{L,R}$ do not
carry any flavor index: all the flavor information is in the
normalization vectors $A$. The specific form of the functions $f,g$
depends on the detail of the bulk physics and will not play any role
for our argument.

The remaining BC's in Eq.~(\ref{eq:BCUV}) determine the masses of
the SM fermions and their KK excitations:

\beq
\begin{array}{c}
f_R (m, z_{UV}) A = m\, g_R (m, z_{UV})\, \mathcal{K}\cdot A \\
\Downarrow\\
\mathcal{K}\cdot A = \frac{f_R (m, z_{UV})}{m\, g_R (m, z_{UV})}\, A
\end{array}
\label{eveq} \eeq This implies that the $A$'s are the eigenvectors
of $\mathcal{K}$ and the eigenvalues of $\mathcal{K}$ will determine
the fermion masses. If the unitary matrix $U$ diagonalizes
$\mathcal{K}$, that is $\mathcal{K} = U \mathcal{K}^{diag}
U^\dagger$, then the normalized eigenvectors are given by
\begin{equation}
A^{\alpha}_{(i )}= U^{\alpha}_i ,
\end{equation}
where the lower index on $A$ indicates which mass eigenstate we are
considering, and the upper index is the index in flavor space. Thus
the U matrix determines in which direction in flavor space the
various mass eigenstates are pointing.  The solutions of
(\ref{eveq}) will thus consist of 3 distinct towers of fermions
(that include the light SM fermions) corresponding to the KK towers
of the three generations. The actual spectrum is then determined by
the equations

\beq
 \frac{f_R (m_i, z_{UV})}{m_i\, g_R (m_i, z_{UV})} = k_i\,, \quad i = 1\dots 3
\eeq where $k_i$ are the eigenvalues of the matrix $\mathcal{K}$.
These equations will determine the masses of the light quarks, and
their KK states.

It is now simple to verify our claims: in the neutral sector, all
the couplings are diagonal. In fact, they will either come from bulk
or IR brane kinetic terms and thus be proportional to $U^\dagger U =
1$, or from the UV brane kinetic terms and thus will be proportional to
$U^\dagger \mathcal{K} U = \mathcal{K}^{diag}$. Let us stress here
that this conclusion can be applied not only to the SM light
particles, but also to fermion and gauge resonances. For the charged
$W$ and its resonances, the couplings are diagonal in flavor space.
However, if the matrices $\mathcal{K}_u$ and $\mathcal{K}_d$ are
misaligned, the couplings to the mass eigenstates will be
proportional to $U^\dagger_u U_d = V_{CKM}$: this defines the CKM
mixing matrix in this scenario. Note that this conclusion can be
applied to not just to the $W$ KK states, but also to extra charged vectors arising from
SU(2)$_R$ since they vanish on the UV brane, so their
couplings are necessarily flavor diagonal.

Finally, the model may also contain extra exotic massive quarks that
can couple to the SM ones via the $W$: in this case, such couplings
may be proportional to a different mixing matrix, for instance
$U_u^\dagger U_{q'} \neq V_{CKM}$. However, their effect is model
dependent and will only enter at loop level.

Let us now count how many mixing parameters and CP violating phases
one has in this setup. We assume that we have $N$
generations,  and we are allowing a separate kinetic mixing matrix
for the right-handed up and down quarks. These kinetic mixing terms
are described by two hermitian $N\times N$ matrices, and therefore
in total there are $2N^2$ real parameters. The parameters of a
general hermitian matrix can be divided into the real diagonal
components, the number of off-diagonal components and the phases
of the off-diagonal components. Thus in total we have $2\times (N+
N(N-1)/2)= N(N+1)$ real parameters, and $N(N-1)$
phases. We are free to make single SU($N$) unitary transformation on both up-type and
down-type quarks simultaneously, since this is an unobservable redefinition of flavor,
that leaves the physics invariant. This SU($N$) symmetry accounts for $N(N-1)/2$ real
parameters and $(N-1)(N+2)/2$ phases. Thus we
are left with $2N+N(N-1)/2$ observable real parameters, however $2N$ of these
correspond to the quark masses. So there will be $N(N-1)/2$ mixing
angles. We are also left with $(N-1)(N-2)/2$ physical
phases. This exactly reproduces the usual CKM picture of CP
violation.

\section{Holographic interpretation in warped space}
\label{sec:CFT} \setcounter{equation}{0} \setcounter{footnote}{0}

The setup used in this paper has a natural four dimensional
explanation in terms of the AdS/CFT correspondence. The 
conjecture is that a 5D theory in AdS space is equivalent to a 4D
conformal field theory. In our  case we are considering a finite
slice of AdS$_5$. The UV (or Planck) brane would correspond to the
CFT having a UV cutoff, and the IR (or TeV) brane to spontaneous
breaking of conformal invariance by strong dynamics. Here we are in
addition requiring that there are some additional {\it global
symmetries} in the 5D theory. This is somewhat unusual, since the
usual lore about AdS/CFT is that a global symmetry of the CFT
corresponds to a {\it gauge} symmetry in the bulk. If we accept that
this is the only reasonable interpretation, we can still make this
bulk gauge symmetry behave almost like a global symmetry by
taking the bulk gauge coupling to be very small.

The CFT interpretation is the following: there is a CFT,
which  has a global symmetry U(3)$_Q\times$U(3)$_{qR}$. This global
symmetry is then spontaneously broken by the CFT interactions that
become strong in the IR (which  is related to the breaking of the
conformal invariance and also of electroweak symmetry) to U(3)$_V$.
The SM fermions are linear combinations of elementary fermions and
of composite states. The elementary fermions do not feel electroweak
symmetry breaking directly, only through the mixing with the
composite modes. The elementary left handed fields respect the same
U(3)$_Q$ flavor symmetry as the conformal sector. However, due to
the misalignment of the kinetic terms of the elementary right-handed
fermions, the U(3)$_{uR}\times$U(3)$_{dR}$ symmetry of the  elementary
sector and the U(3)$_{qR}$ symmetry of the conformal sector are
broken down to U(1)$_{qR}$. Since the CFT also spontaneously breaks
U(3)$_Q\times$U(3)$_{qR}$ to U(3)$_V$ in the end overall there is no
flavor symmetry left unbroken (except of U(1)$_V$ which is
identified with overall baryon number), leading to the possibility
of quark mixing. However, as explained in Sec.~\ref{sec:symmetry}
this global symmetry breaking pattern is sufficient to ensure that
in the neutral current sector there is a
U(1)$_u\times$U(1)$_c\times$U(1)$_t\times$U(1)$_d\times$U(1)$_s\times$U(1)$_b$
symmetry unbroken protecting the theory from FCNC's.

Finally, we need to discuss the fate of the bulk gauge bosons (which
are the consequence of the global symmetry of the CFT). In the CFT
language these will just be a towers of spin 1 modes. As already
discussed, we can reduce the global symmetries of the CFT (i.e. the
gauged symmetries in the bulk) to SU(3)$_Q\times$SU(3)$_{qR}$
without affecting our symmetry argument: this minimal choice ensures
the absence of massless degrees of freedom (like scalar goldstone
bosons and/or massless gauge bosons). Since the elementary sector breaks
SU(3)$_{qR}$, the only gauge symmetry that survives on the UV brane
 is SU(3)$_Q$. Therefore, in addition to the usual
tower of KK states with masses proportional to the IR scale $R'$,
there will be a lighter adjoint of SU(3)$_Q$. The mass of this
flavor gauge bosons is model dependent, but will generically be
suppressed with respect to the IR scale by $\sqrt{\log R'/R}$.
Numerically, it will be roughtly a factor of 10 lighter than the
first KK state, and as low as the $W$ mass in Higgsless models.

One may worry that these new gauge bosons whose masses can be quite
low will themselves mediate flavor changing interactions and
therefore impose an incredibly tight bound on the gauge couplings.
However, as we will show, this is not the case. In fact, the only
source for flavor violation here is the misalignment between the up
and down type quarks. Such misalignment is given by the matrices
$U_u$ and $U_d$ that diagonalize the UV kinetic terms
$\mathcal{K}_u$ and $\mathcal{K}_d$. Therefore, the mass eigenstates
will couple to different combinations of flavor gauge bosons: for
example, in the left-handed sector we have couplings like:

\begin{equation}
i g_Q \bar{u}_\ell^i\gamma_\mu u_\ell^j\, [U^\dagger_u \cdot T^a \cdot U_u]_{ij}\, W_{Q \
\mu}^a, \ \ i g_Q \bar{d}_\ell^i\gamma_\mu d_\ell^j\, [U^\dagger_d \cdot T^a \cdot U_d]_{ij}\,
W_{Q \ \mu}^a\,;
\end{equation}
where $T^a$ are the generators of SU(3)$_Q$. We can immediately see
that no flavor changing operator will involve only up or down type
quarks (which would in fact correspond to highly constrained
FCNC's). Therefore, the only flavor violating 4-fermion operators
must involve the exchange of a $W_Q$ gauge boson between up and down
currents, where the misalignment has a physical effect:
\begin{equation}
(\bar{u}_\ell^i \gamma_\mu u_\ell^j)\, (\bar{d}_\ell^k \gamma^\mu
d_\ell^l)\,.
\end{equation}
The explicit calculation shows that the coefficient of this operator
is given by
\begin{equation}
U_u^{\dagger\ ii'} U_u^{jj'} U_d^{\dagger\ kk'} U_d^{nn'}
\frac{g_Q^2}{M_{W_Q}^2} \sum_a T^a_{i'j'}T^a_{k'n'}
\end{equation}
Using an SU(3) and a Dirac Fierz identity this operator is equivalent to (up to
flavor conserving terms)
\begin{equation}
-\frac{g_Q^2}{2 M_{W_Q}^2} \left[ V_{in} \bar{u}_\ell^i \gamma_\mu
d_\ell^n \right]\left[ V_{kj}^\dagger \bar{d}_\ell^k \gamma_\mu u_\ell^j
\right]
\end{equation}
This is exactly equivalent to the effect of the ordinary W-boson,
with a suppressed gauge coupling $g_Q$ replacing the standard
SU(2)$_L$ coupling $g$. Thus we conclude that the MFV prescription also applies to the
flavor gauge bosons and there is no bound on such operators from flavor 
physics. The actual bounds will come from the traditional electroweak precision 
bounds on flavor conserving operators. The induced four fermi operators 
will be of the form $qqqq$
and thus are not very strongly constrained by  precision electroweak
measurements. A few TeV suppression scale should be sufficient. In
the  SU(3)$_{uR}$, SU(3)$_{dR}$ sector this can be ensured if the
effective 4D gauge coupling is smaller by a factor of 2-3 than the
ordinary SM weak couplings. This can be achieved by choosing
$g_{5,u}\sim g_{5d} \leq 0.3 g_{5L}$. However, as we discussed the
gauge boson corresponding to the
SU(3)$_Q\times$U(1)$_{uR}\times$U(1)$_{dR}$ symmetry can be much lighter
than the others, due to the fact that it is not broken explicitly in
the elementary sector (but only spontaneously by the CFT). Thus its
coupling must be much smaller than that of SU(3)$_{uR}$, SU(3)$_{dR}$. One
may worry that as we take the gauge coupling to zero the mass of
this gauge boson will vanish. However, we know from higgsless models
that this is not the case, and we expect that these gauge bosons
could have a mass at least that of the ordinary $W,Z$ (and in general of order $1/(R'\sqrt{\log R'/R})$). 
To suppress contributions to 4-fermi operators by a few TeV one should make sure
that the gauge coupling in this sector (assuming $1/R' \sim 1$TeV) is at most $g_{5Q} \leq 0.1
g_{5L}$. In this case all 4 fermi operators induced by this very
weakly coupled gauge boson will be negligible.

\section{Applications to models}
\label{sec:models} \setcounter{equation}{0}
\setcounter{footnote}{0}

In this section we show how the general ideas explained above can be applied to obtain concrete 
realistic warped space models with flavor symmetry. We will be focusing on the
RS1 model~\cite{RS1,ADMS} and the Minimal Composite Higgs (MCH)~\cite{MCH} model.
In these models there is a light Higgs localized on or around the TeV
brane with the usual SM VEV of size $\sim 246$ GeV. In both cases unitarization
of WW scattering happens via Higgs exchange, and the KK resonances of the
$W,Z,g$ are heavy $m_{KK}\geq 2$ TeV, corresponding to $1/R'\geq 1$ TeV. The
main difference is that in RS1 there is a little hierarchy problem (i.e. there is
no understanding of why $vR'\ll 1$), while in the MCH model
this is explained since the Higgs is also a pseudo-Goldstone boson of an
approximate global SO(5) symmetry (which also incorporates custodial SU(2)). From the
point of view of fermion representations the main difference is that in RS1 the
bulk fermions are in $(2,1)+(1,2)$ of SU(2)$)_L\times$SU(2)$_R$, while in the
MCH they are in the spinor 4 of SO(5).

\begin{figure}[tb]
\begin{center}
\includegraphics[width=8cm]{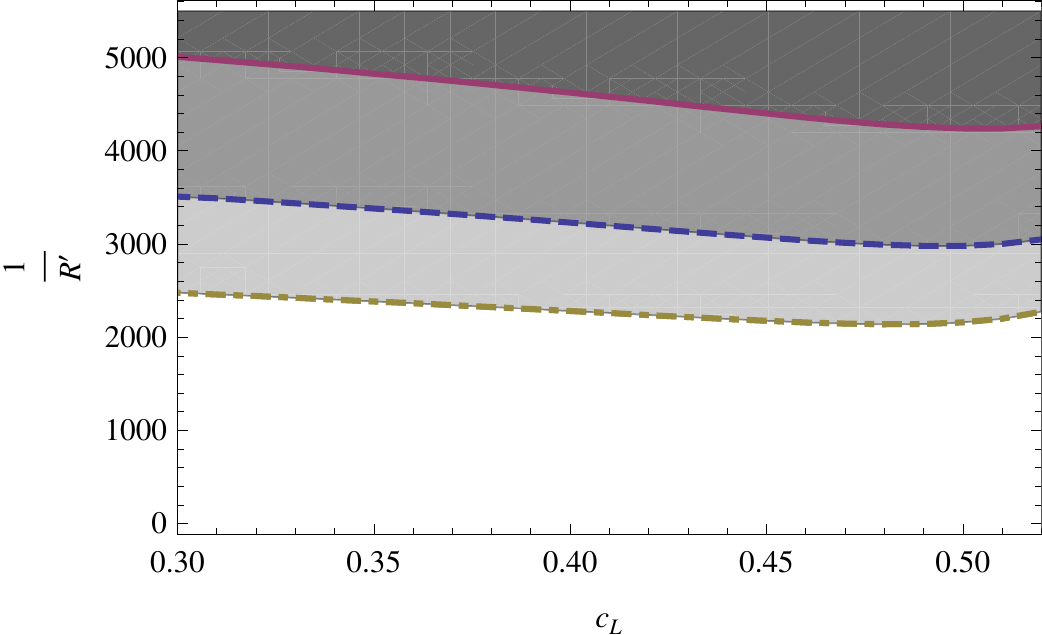}\quad \includegraphics[width=7.5cm]{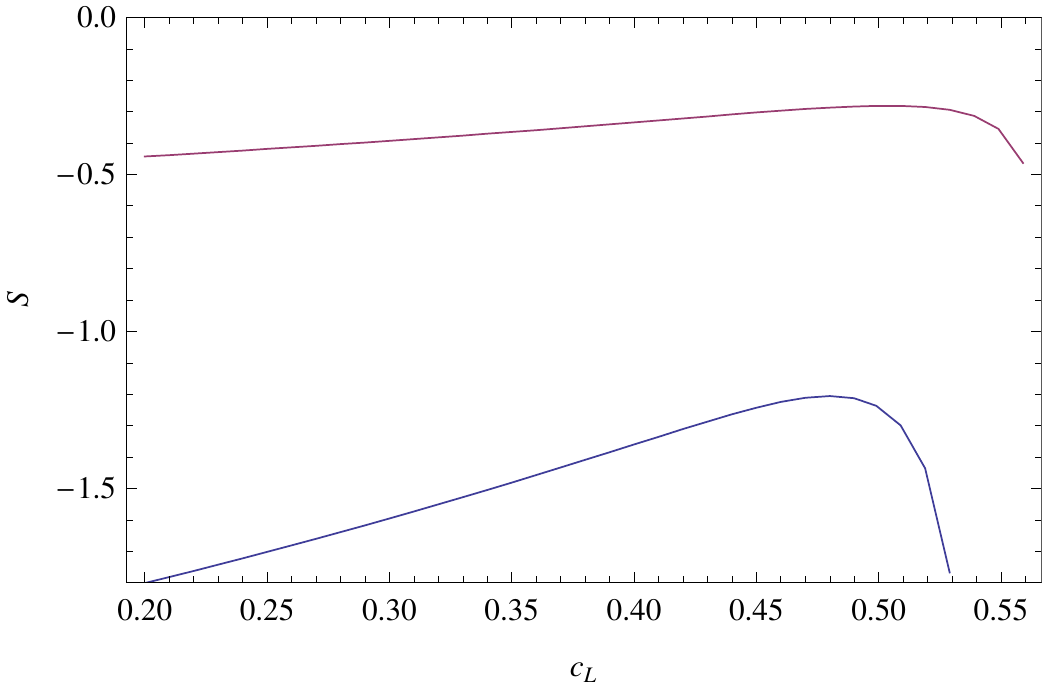}
\end{center}
\caption{Left panel: We show the lower bounds on $1/R'$ as a function of $c_L$ for which $|S| < 1,0.5,0.25$ (bottom to top). Right panel: $S$ in RS1 as a function of the bulk mass $c_L$ for $1/R'= 2$ and $4$ TeV. In both cases we have set $c_R=0$.} \label{S}
\end{figure}

The GIM mechanism outlined in the previous sections can be applied
without any complication to the first two generations of quarks for
both models. The reason is that  if  the heaviest quark mass
is the charm then any overall deviation of fermion wave functions
from those of zero modes will be proportional to $(m_c R')^2 \sim
10^{-6}$. This is no longer true if one also includes the top quark.
Then  $(m_t R')^2 \sim 10^{-2}$, so a percent-level overall shift in
the fermionic gauge couplings would follow. Since these couplings
are measured at the $\sim 0.1\%$ level this would be inconsistent
with the  precision electroweak observables. If all fermions
(including leptons) had a universal $c_L$ and Dirac mass on the TeV
brane, then this could be reinterpreted as a large negative shift in
the effective $S$-parameter, which is unacceptable. This is
illustrated in the right panel of Fig.~\ref{S}, where we show the effective
$S$-parameter as a function of $c_L$ and of $1/R'$ for the RS1 case.
Note that one striking difference of this plot vs. that for
traditional RS1 with bulk fermions is that the $S$-parameter does
not cross through zero around $c_L\sim 1/2$. A simple explanation of
this overall shift in $S$ is the following: the large top mass
usually accompanies an unacceptably large shift in the left handed
couplings of the bottom. By introducing the flavor symmetry we
eliminate the {\it relative} shifts in the couplings, at the price
of introducing an {\it overall} shift that is S. To avoid this we
clearly need to treat the third generation separately. A nice
solution for avoiding a large correction in the $Zb\bar{b}$ coupling
has been proposed by Agashe et al.~\cite{newcustodial}. The main
idea is to use different embeddings of the SM fermions into the
custodial symmetry. The simplest possibility for a single generation
is the following:
\begin{equation}
\begin{array}{cccc}
 & SU(2)_L & SU(2)_R & U(1)_X \\
 Q_L & \Yfund & \Yfund & \frac{2}{3} \vphantom{\sqrt{\frac{2}{3}}}\\
 t_R & 1 & 1 & \frac{2}{3}\vphantom{\sqrt{\frac{2}{3}}} \\
 b_R & 1 & \Ysymm & \frac{2}{3}\vphantom{\sqrt{\frac{2}{3}}} \\
 \end{array}
 \label{newcustodial}
\end{equation}

Zero-modes for the additional fields are projected out using the
BC's. We can then eliminate/reduce FCNC's in two separate ways using
these representations.

\vspace*{0.2cm}

{\bf Model 1.}

\vspace*{0.2cm}

In this setup there are no FCNC's (thus it corresponds to a MFV scenario, i.e. a model
with a built-in GIM mechanism),  no large corrections to the
$Zb\bar{b}$ coupling, and no large vertex corrections (or $S$-parameter). The proposal is to
use the representations in (\ref{newcustodial}) for all three
generations, and to also break the flavor symmetry in the up-sector
from U(3) to U(1)$^3$ by adding different bulk and TeV brane localized masses
for the 3 up-type quarks. This will allow us to avoid a large
overall correction for the light quark couplings,
since the light quark wave functions will not be modified strongly
by the Dirac masses. The use of the unconventional representations
will protect the bottom quark from large vertex corrections (but not
the top or other up-type quarks). The Dirac masses on the TeV brane are assumed to be of the
form (in flavor space)
\begin{equation}
Q_L^s \left( \begin{array}{ccc} m_u \\ & m_c \\ & & m_t \end{array}
\right) t_R + m_b\,  Q_L^t \left( \begin{array}{ccc} 1 \\ & 1 \\ & & 1
\end{array} \right) b_R
\end{equation}
where $Q_L^{s,t}$ are the singlet and triplet components  under SU(2)$_D$ of the bidoublet $Q_L$
and we have inserted a bifundamental Higgs VEV. The key
observation is that the breaking of the up-type flavor symmetry to
U(1)$^3$ does not get communicated into the bottom sector due to the
fact that $t_R$ is a singlet under the custodial symmetry and thus
this mass term does not involve any quarks with the quantum numbers
of the bottom that could potentially mix with the bottom (this is
unlike the case using the standard representations where the $(t_R,b_R)$ are in a doublet of SU(2)$_R$). 
Thus
in the NC sector the full flavor symmetry in the down-type quark
sector is unbroken on the TeV brane. Then if we introduce all flavor
mixing via UV brane localized kinetic mixing terms for the down-type
quarks only, all FCNC's will be avoided, so that this setup also has
a GIM mechanism. 

\begin{figure}[tb]
\begin{center}
\includegraphics[width=10cm]{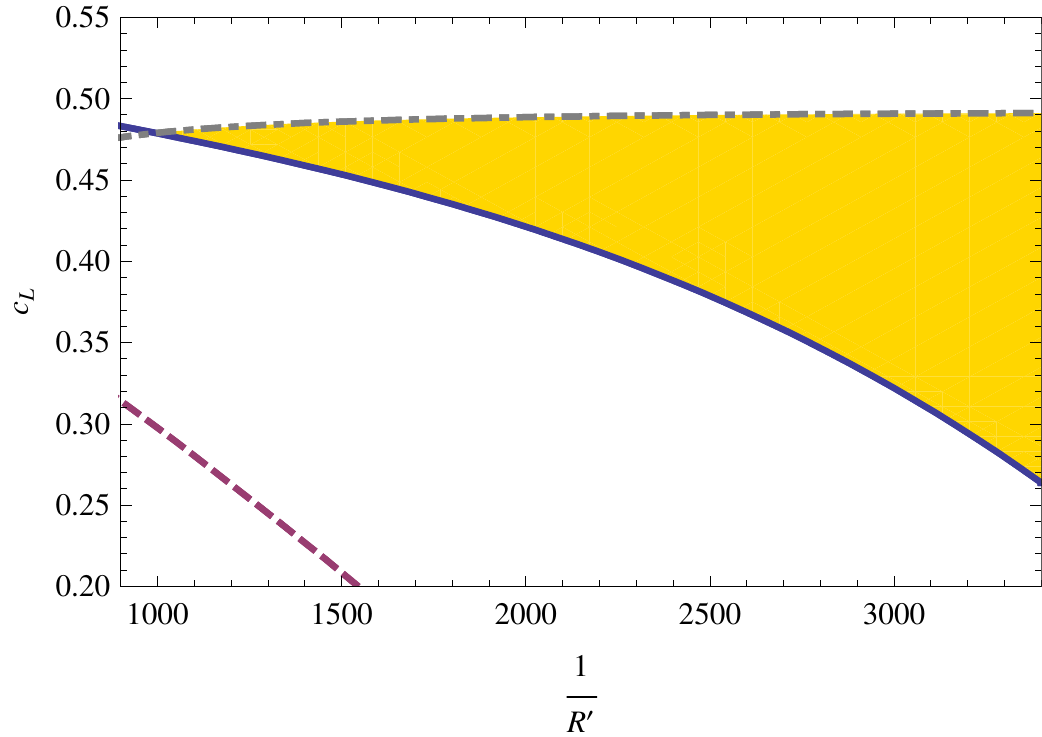}
\end{center}
\caption{The allowed region of the parameter space in Model 1 with MFV. The upper dash-dotted curve shows the 
perturbativity bound on the parameter $c_L$ assuming that the right handed top is strongly localized on the 
TeV brane (we chose $c_R=1$, and required that the Yukawa coupling is less than four). The middle
solid curve shows the region where the deviation of the coupling of the left handed up-type quark is 
below 0.5\%. The allowed region is the shaded one between these two curves. The lowest dashed curve shows 
the bound from the coupling of the KK mode of the $Z$ coupling to light left handed quarks. Such a coupling would
generically induce four fermi operators involving leptons. The bound obtained in this plot is requiring that
the $q/M_{Z'}$ ratio of coupling to mass is less than 1/(5 TeV)~\cite{CCMS}.
We can see that this bound is irrelevant for the allowed region.} \label{newcustodianPlot}
\end{figure}

What needs to be checked is that this setup is also consistent with
other  precision electroweak constraints. In particular the main worry is
that since we are still requiring an SU(3)$_L$ flavor symmetry for the
left handed quarks the requirement for a large top mass may still 
result in a large vertex corrections. Generic electroweak precision bounds on 
models using the new representations under custodial symmetry were presented in~\cite{newcustodialbounds}.
Note however, that the choice of the new representations
under the custodial symmetry is designed such that the down-type left handed quarks
do not get much of a vertex correction, while there is no protection mechanism for the up-type quarks.
Therefore by construction the corrections are non-oblique and one should not use the
the oblique parameter formalism to estimate the sizes of the corrections. 
A simple scheme~\cite{CCMT} to estimate the bounds from the  precision electroweak observables 
is to simply fix the $W,Z$ masses and 
the electromagnetic couplings as the input parameters, and shift all corrections 
into vertex corrections to $Z$ and $W$ couplings. The leptons are not a problem: they can be either in the bulk or localized near the Planck brane, depending on how low $1/R'$ is. Concerning the quarks, right handed up and down-type quarks are near the Planck brane (except for the top, whose couplings are not strongly constrained), thus the deviation of their couplings to the $Z$ is under control. Since there is an explicit protection mechanism
for the left handed down-type quarks, the only potentially dangerous corrections are those to the light left handed up-type 
quarks. To find the experimental bounds on this model from this effect we have first calculated the 
maximal value of $c_L$ for which a sufficiently heavy top
mass can still be obtained for perturbative values of the Yukawa 
couplings. This bound turns out to be relatively insensitive to the values of
$R'$ and the bound is around $c_L \leq 0.47$. However, this bound is somewhat sensitive to the 
localization parameter $c_R$ of the right handed top. What one then needs to check is
the correction to the vertex corrections to $g_{Z{u_L}{u_L}}$ are not too large. Note, that 
the main constraints on this coupling come from the measurements involving hadronic final states at 
LEP, for example $\Gamma (Z)$ and $\sigma ({\rm hadron})$. Using the method of~\cite{CCMS} we estimate that 
a reasonable 3 sigma bound on the deviation of this coupling is about $\pm 0.4$ percent. In addition we also 
need to make sure that the couplings to the KK $Z^\prime$ bosons or KK gluons will not be too large to generate
flavor invariant 4 fermi operators. Our results are summarized in Fig.~\ref{newcustodianPlot}
where we show that unless
$1/R'$ is very small the shift in $g_{Z{u_L}{u_L}}$ is acceptably small,
while the coupling of the light fermions to the KK $Z^\prime$ always remain within an acceptable range. A bound 
on $1/R'$ of order $\sim 1 - 1.5$ TeV follows from these constraints. In Table~\ref{table:example} 
we show an example for the shifts in the couplings for an allowed point in the parameter space 
for a low value of $1/R'$. Another electroweak precision bound that one may consider (depending on the exact 
treatment of the right handed up quarks) are four fermi operators of the form $eeu_Ru_R$. These will be generated
if the bulk SU(3)$_{u_R}$ flavor symmetry is maintained (i.e. we use the same $c_R$ for all $u_R$'s). In this case the 
light $u_R$'s will have an enhanced coupling to the KK $Z'$ mode. This operator is only constrained from the 
LEP2 $e^+e^-\to q\bar{q}$ cross section measurement and its coefficient is not very strongly constrained. Also, the enhancement
of the $u_R$ couplings is partly offset by the suppression of the electron couplings due to them being localized on 
the UV brane. Using the $\chi^2$ provided in~\cite{Witek} we have checked that these operators do not further reduce
the allowed parameter space in Fig.~\ref{newcustodianPlot}.

\begin{table}[th]
\begin{center}
\begin{tabular}{|l||l|l|l|l|}
\hline
u & $\gamma_L^u = -3.1   $ & $\omega_L = -0.48$ & $\gamma_R^u =0.76  $ & $\omega_R < 10^{-7}$ \\
d & $\gamma_L^d = 1.4  $ &                     & $\gamma_R^d =-0.012 $ & \\
\hline
c & $\gamma_L^c = -3.1  $ &$\omega_L=-0.48$  & $\gamma_R^c =0.76  $ & $\omega_R  < 10^{-3}$ \\
s & $\gamma_L^s = 1.4 $ &                 & $\gamma_R^s =-0.016$      &  \\
\hline
t & $\gamma_L^t = -3.9 $ & $\omega_L=-0.85$ & $\gamma_R^t = 20 $ & $\omega_R = -2.2 $\\
b & $\gamma_L^b =  1.4  $ &               & $\gamma_R^b =-7.1$    &  \\
\hline

\end{tabular}
\caption{
Per mille relative deviations of the effective couplings to the SM values of the fermion gauge coupling strengths 
for a particular allowed point from Fig.~\ref{newcustodianPlot},
 chosen to correspond to $1/R'=1.5$ TeV, $c_L =0.47$, $c_R=-0.51$, 
and $c_{tR}=1$. We parameterize the deviation by
$g^Z_{f_L} = ( 1+\gamma_L^f ) \frac{g}{\cos \theta_W} (
T_3 - \sin^2 \theta_W\, Q )\,$, $ g^W_{f_L} =(
1+\omega_L^f ) g\,$ and similarly for the right-handed couplings.
}\label{table:example}
\end{center} \end{table}

Next we can check that the number of flavor parameters really agree with that expected from the 
ordinary CKM mechanism. The mixing is described now by a single hermitian kinetic 
mixing matrix for the right handed down quarks, which is parametrized by 
$N(N+1)/2$ real numbers (angles) and $N(N-1)/2$ phases. Of these real numbers $N$ correspond
to the quark masses in the down sector, and we can still use the U(1)$^N$ unbroken 
flavor symmetry to remove $N-1$ phases (one overall is baryon number). Thus we 
again end up with $N(N-1)/2$ mixing angles and $(N-1)(N-2)/2$ CP violating phases
as expected.

Finally, we comment on the possible presence of familons and alignment issues in this model. The symmetry breaking 
pattern is the following. On the UV brane we have and SU(3)$_L\times$SU(3)$_{u_R}$ symmetry, in the 
bulk we have SU(3)$_L\times$SU(3)$_{d_R}\times$U(1)$_{u_R}^3$ while on the TeV brane just 
U(1)$_{diag}^3$. Since the coupling of the $u_R$ quarks turns out to be quite insensitive to
the value of $c_R$, we are not necessarily forced to break the SU(3)$_{u_R}$ symmetry in the 
bulk to U(1)$_{u_R}^3$. In the case of a bulk SU(3)$_{u_R}$ symmetry we would not have to 
deal with the question of why the U(1)$^3$ symmetries in the bulk and the brane are aligned. If we do 
break the symmetry in the bulk, we have to insist that the same spurion is used to break the bulk and the 
brane symmetries. 

In the CFT interpretation there is a gauged SU(3)$_L\times$U(1)$_{u_R}^3$ symmetry (or 
SU(3)$_L\times$ SU(3)$_R$) and
an additional SU(3)$_{d_R}$ global symmetry, all of which are broken by the CFT to a U(1)$_{diag}^3$ global symmetry. Thus 
there would be familons corresponding to the breaking of SU(3)$_{d_R}$. In order to make these
appropriately heavy there has to be an additional explicit breaking of this symmetry, which in the 5D picture
corresponds to a bulk Higgs for SU(3)$_{d_R}$ which is however not coupled to the bulk fermion fields.
Such a bulk Higgs can also make bulk gauge bosons arbitrarily heavy and effectively decouple them while leaving
a global symmetry in the fermion sector. 


\vspace*{0.2cm}

{\bf Model 2.}

\vspace*{0.2cm}

In this scenario, we treat the third generation differently from the two light generations, and therefore we only impose a SU(2)$\times$U(1) flavor symmetry.
The symmetry breaking pattern will be the same as in Sec.~\ref{sec:symmetry} with SU(3) replaced by this reduced symmetry.
The new feature of this scenario is that the third generation can have different bulk and IR masses, and even be in different representations of the bulk gauge symmetries.
We will  leave the light generations in the usual representations and localized towards the UV brane, while the third 
generation is in the new custodial representation and can be localized near the IR brane.
As a result we will clearly not have a
problem with a large shift in the $S$-parameter and the deviation of
the $Zb\bar{b}$ coupling will be sufficiently reduced due to the use
of the representations (\ref{newcustodial}).
The drawback of this scenario is that FCNC's will be generated, not only involving the top and bottom, but also for the light generations.
However, due to the SU(2) symmetry, those FCNC's will have a particular form and may result in weaker bounds than in the usual RS1 case~\cite{Gustavo, warpedconstraints}.

Let us first note that due to the heaviness of the top quark, the right handed top should be 
strongly localized on the TeV brane: therefore the effective kinetic mixing terms in
the right handed up-sector only involve the first two generations
(since the wave function of the right handed top is negligible on
the UV brane). This matrix can be diagonalized using the SU(2) symmetry, so that all the off-diagonal terms are in the down sector.
As a consequence, FCNCs in the up-quark sector are negligible.

In the down sector the situation is more complicated, since the R bottom is also localized towards the UV brane and we do need to generate a mixing with the third generation. 
The difference with respect to the case analysed in Sec.~\ref{sec:symmetry} is that the wave function of the third generation is different from the wave function of the light generations.
This means that the elements of the mixing matrix $\mathcal{K}$ will acquire different coefficients, depending on the quantity we are interested in: for instance, the matrices entering the boundary conditions and the couplings of the right-handed down quarks to the $Z$ will be different.
However, due to the SU(2) symmetry, the effect of the different wave functions will preserve a block structure as follows:
\beq \left( \begin{array}{cc}
a \mathcal{K}_{(12)} & b \mathcal{K}_{(13)} \\
b \mathcal{K}^T_{(13)} & c \mathcal{K}_{33}
\end{array} \right)
\eeq 
where $\mathcal{K}_{(12)},\mathcal{K}_{(13)},\mathcal{K}_{33}$ are the appropriate blocks of
the UV kinetic mixing matrix, and the coefficients $a,b,c$ depend
on the details of the wave functions and the quantity we are calculating. 
If we diagonalize the $2\times 2$ block $\mathcal{K}_{(12)}$ with an SU(2) rotation, we only generate the Cabibbo angle in the couplings of the $W$, 
but no FCNC in the couplings with neutral gauge bosons, as in Sec.~\ref{sec:symmetry}.
We still have off-diagonal terms proportional to $\mathcal{K}_{(13)}$ that cannot be diagonalized in the same way due to the effect of the different wave functions.
However, in this basis every flavor changing effect involving the
down quark must be proportional to the top component of the vector
$\mathcal{K}_{(13)}$ since this is the only flavor violating matrix element
involving the down quark. The CKM matrix connecting the
down to the third generation must also be directly proportional to
this matrix element, since there is no flavor mixing involving the
top directly. Applying a similar argument for the strange quark we
find that the flavor changing couplings must always be suppressed by
the appropriate CKM matrix elements:
\beq \label{eq:pattern}
g_{Zds} &\sim& V_{td} V_{ts} \delta\\
g_{Zdb} &\sim& V_{td} \delta \\
g_{Zsb} &\sim& V_{ts} \delta \eeq
This pattern will show up for the couplings of the $Z$ as well as other neutral massive bosons, and for both L and R fermions. 
This means that this model is a simple implementation of  NMFV,
where all flavor violation is due to the flavor violating interactions with the third generation.
The main difference compared to the traditional RS models is that here the NMFV structure of the corrections (by which we mean that all additional flavour 
violation must proceed through the third generation) is due to symmetries (rather
than smallness of wave function overlaps) and thus here there will never be any direct contributions to $K\bar{K}$ mixing that are 
not proportional to the CKM matrix elements involving the third generation.

We can also easily understand the origin of the parameter $\delta$. There are two sources for gauge boson couplings:
bulk kinetic terms and the brane localized kinetic mixing terms. The bulk kinetic terms are diagonal
but not necessarily universal due to the different wave function for the bottom, thus take the 
form diag$(g_1,g_1,g_3)$ in flavor space before diagonalizing the localized mixing terms. The localized mixing
terms have an arbitrary structure in the down sector, but the main point is that once the kinetic terms are 
diagonalized so will the resulting contributions to the neutral gauge boson couplings as well. Therefore, the 
only source of off-diagonal couplings is the difference $\delta \propto g_3-g_1$. For the case of left handed fields 
there is no localized kinetic term so $g_3-g_1$ is directly proportional to $g_{Z{b_L}{b_L}} - g_{Z{d_L}{d_L}}$. However
for the right handed fields there can also be a contribution to $\delta$ from hierarchies in ${\cal K}_d$. Thus the precise value
of $\delta$ for the right handed fields is likely to be quite model dependent.

Now we can estimate the bounds on the model using the above structure for the corrections.
For simplicity we
will apply a bound of the form~\cite{Zprime} 
\beq \frac{g_{Zds}}{g}
\frac{m_Z}{m_{Z^\prime}}  < 10^{-5} \eeq 
where $Z^\prime$ stands for an arbitrary neutral gauge boson in the model.

If we apply this for the $Z$ in the
left handed sector we find that the requirement is 
\beq
\frac{\delta_{Zbb}}{g} < 3\% 
\eeq 
which is weaker than the
direct bound from  precision electroweak observables. For the effects
of the $Z^\prime$ in the L sector we find (assuming that there is
not much suppression in the coupling of the light L quarks
to the $Z^\prime$ compared to the SM value)
\begin{equation}
m_{Z^\prime} > 10^5\, V_{td} V_{ts} \, m_Z\, \sim (3\, \mbox{TeV})
\end{equation}
A slightly stronger bound would be obtained by considering the KK mode of the gluon field. 
For the right-handed quarks, the presence of non-diagonal kinetic terms makes the analysis more involved: 
however, we can observe that all the right-handed down quarks are very localized on the UV brane, 
so that their couplings to the KK gauge bosons are suppressed by a log factor compared to the SM ones:

\begin{equation}
\delta \sim \frac{g}{\sqrt{\log R'/R}}
\end{equation}
The bound on the mass of the $Z^\prime$ is therefore weaker than from the
L sector considered above. However, we do not have a reliable
estimate for the suppression of the flavor violating right handed
couplings of the ordinary $Z$, which may be still be a problem and
needs to be calculated numerically.

We can see one additional advantage of this setup versus the traditional split-fermion approach
to RS flavor, and that is when considering the additional CP phases. In the traditional setup
there is a CP problem, i.e. there are additional physical CP phases that have no reason to be
small, and which will tighten the bounds on the KK scale to $\sim 8$ TeV~\cite{NMFVbound}. However,
in this setup the symmetries will forbid the appearance of an additional physical CP 
phase. As we have seen in this setup all flavor violating parameters originate from the 
mixing matrices ${\cal K}_{u,d}$.\footnote{The bulk mass parameters are real, while the phases of the IR brane 
Yukawa couplings can be absorbed into the fermion fields without affecting the mixing matrices on the UV brane.} 
For the model under consideration ${\cal K}_d$ is a generic 
$3 \times 3$ hermitian matrix and ${\cal K}_u$ a generic $2 \times 2$ hermitian matrix. Thus ${\cal K}_d$ 
contains 3 complex phases, while ${\cal K}_u$ contains 1. However, the remaining symmetry of the bulk +
IR brane sector is SU(2)$\times$U(1)$\times$U(1)$_B$. The U(1)$_B$ corresponding to baryon number has no effect
on the phases in ${\cal K}_{u,d}$, however the remaining SU(2)$\times$U(1) can still be used to eliminate 
non-physical phases. Since SU(2)$\times$U(1) contains 3 phases, we conclude that in total there is just a 
single CP violating phase in this setup, which should be identified with the CP phase from the 
CKM mechanism. Thus there is no possibility for an additional CP problem to emerge here.

\section{Conclusions}
\label{sec:conclusions} \setcounter{equation}{0}
\setcounter{footnote}{0}

In this work we have explored how to implement an alternative realization of the SM flavor
structure in warped extra dimensional models. The accepted approach is to use the split fermion
scenario where hierarchies are obtained from overlaps of wave functions. Here we 
asked how a more traditional picture based on flavor symmetries can be implemented. We found that 
if there is a sufficiently large bulk+IR brane flavor symmetry, a GIM mechanism can be 
incorporated preventing the generation of FCNC's. In this case all the mixing is obatined 
from UV brane localized kinetic mixings. Inclusion of a large top mass (together with 
electroweak precision constraints) forces us to modify this minimal setup. Models with MFV
can be obtained by putting all SM quarks into new representations under the custodial SU(2) 
symmetry, while models with NMFV are obtained by using the new representations only for the 
third generation quarks.

\section*{Acknowledgments}

We thank Kaustubh Agashe, Yuval Grossman and Gilad Perez for useful discussions
and comments. The research of C.C. and A.W. is supported in part by
the NSF grant PHY-0355005 and in part by the DOE OJI grant
DE-FG02-01ER41206. A.W. is grateful to the Aspen
Center for Physics, where portions of this work were completed. 
The research of G.C., J.G., G.M. and J.T. is 
supported in part by the US Department of Energy under contract
DE-FG03-91ER40674.


\begin{thebibliography}{99}

\bibitem{AS}
N.~Arkani-Hamed and M.~Schmaltz,
  Phys.\ Rev.\  D {\bf 61}, 033005 (2000)
  [arXiv:hep-ph/9903417].


\bibitem{RS1}
L.~Randall and R.~Sundrum,
 Phys.\ Rev.\ Lett.\  {\bf 83}, 3370 (1999)
 [arXiv:hep-ph/9905221].

\bibitem{warpedflavor}
Y.~Grossman and M.~Neubert,
  Phys.\ Lett.\  B {\bf 474}, 361 (2000)
  [arXiv:hep-ph/9912408];
T.~Gherghetta and A.~Pomarol,
  Nucl.\ Phys.\  B {\bf 586}, 141 (2000)
  [arXiv:hep-ph/0003129].

\bibitem{HuberShafi}
 S.~J.~Huber and Q.~Shafi,
  Phys.\ Lett.\  B {\bf 498}, 256 (2001)
  [arXiv:hep-ph/0010195];
 S.~J.~Huber,
  Nucl.\ Phys.\  B {\bf 666}, 269 (2003)
  [arXiv:hep-ph/0303183].


\bibitem{CKM}
N.~Cabibbo,
{\em Phys.\ Rev.\ Lett.}\ 10 (1963) 531;
M.~Kobayashi and T.~Maskawa,
{\em Prog.\ Theor.\ Phys.}\ 49 (1973) 652.


\bibitem{Gustavo}
G.~Burdman,
  Phys.\ Rev.\  D {\bf 66} (2002) 076003
  [arXiv:hep-ph/0205329];
G.~Burdman,
 Phys.\ Lett.\  B {\bf 590}, 86 (2004)
 [arXiv:hep-ph/0310144].

\bibitem{warpedconstraints}
K.~Agashe, G.~Perez and A.~Soni,
  Phys.\ Rev.\ Lett.\  {\bf 93}, 201804 (2004)
  [arXiv:hep-ph/0406101];
  Phys.\ Rev.\  D {\bf 71}, 016002 (2005)
  [arXiv:hep-ph/0408134].


\bibitem{NMFV}
K.~Agashe, M.~Papucci, G.~Perez and D.~Pirjol,
 arXiv:hep-ph/0509117;
Z.~Ligeti, M.~Papucci and G.~Perez,
 Phys.\ Rev.\ Lett.\  {\bf 97}, 101801 (2006)
 [arXiv:hep-ph/0604112].

\bibitem{NMFVbound}

M.~Bona {\it et al.}  [UTfit Collaboration],
  arXiv:0707.0636 [hep-ph];
K.~Agashe {\it et al.},
  arXiv:0709.0007 [hep-ph].


\bibitem{AdSCFT}
 J.~M.~Maldacena,
  Adv.\ Theor.\ Math.\ Phys.\  {\bf 2}, 231 (1998)
  [Int.\ J.\ Theor.\ Phys.\  {\bf 38}, 1113 (1999)]
  [arXiv:hep-th/9711200];
O.~Aharony, S.~S.~Gubser, J.~M.~Maldacena, H.~Ooguri and Y.~Oz,
  Phys.\ Rept.\  {\bf 323}, 183 (2000)
  [arXiv:hep-th/9905111].

\bibitem{AdSCFTpheno}
N.~Arkani-Hamed, M.~Porrati and L.~Randall,
  JHEP {\bf 0108}, 017 (2001)
  [arXiv:hep-th/0012148];
R.~Rattazzi and A.~Zaffaroni,
  JHEP {\bf 0104}, 021 (2001)
  [arXiv:hep-th/0012248].

\bibitem{higgsless}
C.~Cs\'aki, C.~Grojean, H.~Murayama, L.~Pilo and J.~Terning,
  Phys.\ Rev.\  D {\bf 69}, 055006 (2004)
  [arXiv:hep-ph/0305237];
C.~Cs\'aki, C.~Grojean, L.~Pilo and J.~Terning,
  Phys.\ Rev.\ Lett.\  {\bf 92}, 101802 (2004)
  [arXiv:hep-ph/0308038].

\bibitem{GIM}
S.~L.~Glashow, J.~Iliopoulos, and L.~Maiani,
{\em Phys.\ Rev.\ D} 2 (1970) 1285;
N.~Cabibbo,
Phys.\ Rev.\ Lett.\  {\bf 10}, 531 (1963);
M.~Kobayashi and T.~Maskawa,
Prog.\ Theor.\ Phys.\  {\bf 49}, 652 (1973).


\bibitem{tcgim}
R.~S.~Chivukula and H.~Georgi,
Phys.\ Lett.\  B {\bf 188}, 99 (1987);
L.~Randall,
  Nucl.\ Phys.\  B {\bf 403}, 122 (1993)
  [arXiv:hep-ph/9210231].

\bibitem{UEDGIM}
A.~J.~Buras, M.~Spranger and A.~Weiler,
  Nucl.\ Phys.\  B {\bf 660}, 225 (2003)
  [arXiv:hep-ph/0212143];
  A.~J.~Buras {\it et al}.,
  Nucl.\ Phys.\  B {\bf 678}, 455 (2004)
  [arXiv:hep-ph/0306158];
U.~Haisch and A.~Weiler,
  Phys.\ Rev.\  D {\bf 76}, 034014 (2007)
  [arXiv:hep-ph/0703064].


\bibitem{MFV}
E.~Gabrielli and G.~F.~Giudice,
Nucl.\ Phys.\ B {\bf 433}, 3 (1995)
[Erratum-ibid.\ B {\bf 507}, 549 (1997)]
[arXiv:hep-lat/9407029];
A.~Ali and D.~London,
Eur.\ Phys.\ J.\ C {\bf 9}, 687 (1999)
[arXiv:hep-ph/9903535];
A.~J.~Buras {\it et al}., 
Phys.\ Lett.\ B {\bf 500}, 161 (2001)
[arXiv:hep-ph/0007085];
G.~F.~Giudice, G.~Isidori and A.~Strumia,
Nucl.\ Phys.\ B {\bf 645}, 155 (2002)
[arXiv:hep-ph/0207036];
C.~Bobeth {\it et al}., 
Nucl.\ Phys.\ B {\bf 726}, 252 (2005)
[arXiv:hep-ph/0505110];
U.~Haisch and A.~Weiler,
 arXiv:0706.2054 [hep-ph].

\bibitem{newcustodial}
K.~Agashe, R.~Contino, L.~Da Rold and A.~Pomarol,
 Phys.\ Lett.\  B {\bf 641}, 62 (2006)
 [arXiv:hep-ph/0605341].

\bibitem{ADMS}
K.~Agashe, A.~Delgado, M.~J.~May and R.~Sundrum,
  JHEP {\bf 0308}, 050 (2003)
  [arXiv:hep-ph/0308036].

\bibitem{MCH}
K.~Agashe, R.~Contino and A.~Pomarol,
  Nucl.\ Phys.\  B {\bf 719}, 165 (2005)
  [arXiv:hep-ph/0412089];
K.~Agashe and R.~Contino,
Nucl.\ Phys.\  B {\bf 742}, 59 (2006)
[arXiv:hep-ph/0510164];


\bibitem{newcustodialbounds}
M.~S.~Carena, E.~Ponton, J.~Santiago and C.~E.~M.~Wagner,
  Nucl.\ Phys.\  B {\bf 759}, 202 (2006)
  [arXiv:hep-ph/0607106];
  arXiv:hep-ph/0701055.

\bibitem{CCMT}
G.~Cacciapaglia, C.~Cs\'aki, G.~Marandella and J.~Terning,
 Phys.\ Rev.\  D {\bf 75}, 015003 (2007)
 [arXiv:hep-ph/0607146].


\bibitem{CCMS}
G.~Cacciapaglia, C.~Cs\'aki, G.~Marandella and A.~Strumia,
  Phys.\ Rev.\  D {\bf 74}, 033011 (2006)
  [arXiv:hep-ph/0604111].

\bibitem{Witek}
Z.~Han and W.~Skiba,
  Phys.\ Rev.\  D {\bf 71}, 075009 (2005)
  [arXiv:hep-ph/0412166].

\bibitem{Zprime}
P.~Langacker and M.~Plumacher,
  Phys.\ Rev.\  D {\bf 62}, 013006 (2000)
  [arXiv:hep-ph/0001204].

\end{thebibliography}
\end{document}